\documentclass[aps,showpacs]{revtex4}
\usepackage{graphics}
\usepackage{latexsym}
\usepackage[dvips]{graphicx,epsfig}
\usepackage{psfrag}
\newcommand{\be}{\begin{equation}}
\newcommand{\ee}{\end{equation}}
\def\bq{\begin{eqnarray}}
\def\eq{\end{eqnarray}}

\begin{document}
%
\large  
\title{A MODEL FOR QUARK-GLUON PLASMA WITH PENTAQUARK BARYONS AND TETRAQUARK MESONS}
\pacs{ 12.38.Mh ; 12.39.Mk ; 13.40.Gp }
\author{R. S. Kaushal  \\ Department of Physics, Ramjas College (University Enclave), \\University of Delhi, Delhi-110007, India \\ and \\Department of Physics \& Astrophysics, University of Delhi,\\ Delhi-110007, India}

\email{rkaushal@physics.du.ac.in}

\begin{abstract}
With a view to exploring a new kind of phase transition in the process of hadronization of quark-gluon plasma (QGP) we investigate the occurrence of pentaquark baryons and tetraquark mesons in the system. For this purpose, the frame work of an analoguous Saha's ionization formula for the colored ions in the system is used. The study of color-ionic-fraction (CIF) of multiply (color) ionized to unionized quark clusters (termed as "quarkons") as a function of temperature is carried out. It is pointed out that not only the temperature of the fire-ball in the relativistic heavy ion collisions evolves with respect to space and time but also the CIF associated with a particular stage of ionization. Further, for the case of single color-ionization a correspondence of the present results with those available for the bubble nucleation mechanism in QGP is demonstrated.
\end{abstract}

\maketitle

\newpage

\section{Introduction}
It is now well known that heavy ion collision experiments performed at ultra-relativisitc energies ($\geq 200$ GeV/nucleon) can offer the same conditions as they existed at the time of origin of early universe, i.e., the conditions immediately after the big-bang. As a matter of fact the universe, during its space time evolution, has passed through a stage (though for a short period) in which a deconfined state of quarks and gluons (in brief called quark-gluon plasma or QGP characterized as a 'fire-ball') existed and the same is now expected to be formed during mini-bang, i.e., in the relativistic heavy ion collision (RHIC) experiments. In recent years, these studies have been the subject of great interest [1,2]. Particularly, the efforts have been there to look for the underlying mechanism and the processes which have led to the formation of hadronic state of matter out of the deconfined state or the vice versa. Various models to this effect have been discussed in the literature[3]. Until recently the picture considered was as follows: It is believed that initially existing deconfined state of quarks and gluons cooled down and as it evolved with space and time, the formation of hadrons (tri-quark baryons and two-quark mesons) took place via several stages of phase transitions. In particular, this quark-gluon state  first converted into quark gluon soup[4,5], then into the diquark gluon plasma[6] (since diquark is also a colored object) and finally into the hadronic phase.
\par
	It may be mentioned that beside the baryons ($qqq$) and mesons ($q\bar{q}$) the existence of other (likely to be) stable quark clusters (now onward termed as 'quarkons') has been noticed only recently in various experiments[7]. Not only this, several attempts have already been made[8,9] to understand them in theoretical terms. While the theoretical understanding of these newly discovered objects (which in particular, are the pentaquark baryons like $\theta^{+}(1540)$, $\zeta^{0}(3099)$ and $\Xi^{--}(1860)$ and some of them are designated as $(udud\bar{s})$, $(udud\bar{c})$, $(dsds\bar{u})$ etc. and tetraquark mesons like $(ud\bar{u}\bar{d})$) has yet to be established, their role in the physics of QGP has already become[10] of great interst. As a matter of fact the possibility of production of pentaquark $\theta^{+}$- baryon in the RHIC experiments has recently been discussed by Chen et al.[10] in a kinetic model via the processes $KN$$\leftrightarrow$ $\theta$, $KN$ $\leftrightarrow$ $\pi$$\theta$,and $\pi$$N$ $\leftrightarrow$$\bar{K}\theta$. 
\par
	In fact once the existence of pentaquark baryons and tetraquark mesons is confirmed, the studies of QGP would require another stage of phase transition. As a result a new picture of QGP formation in RHIC experiments would emerge. The purpose of the present paper is to discuss one such model of QGP formation, of course, within the frame work of somewhat less refined scheme, i.e., within the frame work of color-quark chemistry which once played[11] an important role in understanding the quark dynamics of hadrons. For this purpose, we shall explore here (perhaps for the first time to the best of our knowledge) the tools of Saha's ionization formula[12] (SIF) but now for color-charges in stead of the customary Coulomb charges. In particular, a computation of color ionic fraction (CIF) will be carried out for different quarkons and 'diquarkons'(diquark clusters).

\par
	In the next section, we review the knowledge of SIF for coulombic ions and discuss the possible extension of results to the case of colored ions. In Section 3, we highlight some typical processes responsible for the occurrence of pentaquark and tetraquark systems and carry out the computation of Saha's color-ionic fraction. A possible connection between the single color-ionization process and the well-studied nucleation mechanism for the hadronic phase transitions, is highlighted, perhaps for the first time, in Section 4. Finally concluding remarks are made in Section 5.
 
\section{Brief Review of Saha (Coulomb) Ionization Formula (SIF)}
A detailed study of ionization of gases by thermal excitation was carried out by M. N. Saha about eighty years ago[12]. Particularly, the question as to what happens when a gaseous mass consisting of atoms is heated to a very high temperature was discussed by Saha and a formula for computing the fraction of the number of $(r + 1)$-times ionized to the $r$-times ionized atoms was derived in terms of pressure, temperature and internal energy of the system.

\par
	The generalized version of the SIF for the chemical reaction $\sum_{r} a_{r}A_{r}$
 $\rightleftharpoons$ $\sum_{s} b_{s}B_{s}$, as an outcome of classical statistics and the law of mass action provides, (cf. Ref.(12), p. 658)

\begin{equation}
ln K_{P} = -\frac{U}{RT} + \frac{5}{2} ln T + ln I + ln Z,
\end{equation}
where the fractions,$K_{P}$, $I$, and $Z$ of corresponding partial pressures $(P's)$, intrinsic constants $(I's)$, and parition functions $(Z's)$ for products and reactants in the given chemical reaction, are defined as

\begin{equation} 
 K_{P} = \frac{\prod_{s} P_{B_{s}}^{b_{s}}}{\prod_{r}P_{A_{r}}^{a_{r}}}; I = \frac{\prod_{s} i_{B_{s}}^{b_{s}}}{\prod_{r} i_{A_{r}}^{a_{r}}}; Z = \frac{\prod_{B_{s}}(Z_{r}Z_{\nu}Z_{e}...)^{b_{s}}}{\prod_{A_{r}}(Z_{r}Z_{\nu}Z_{e}...)^{a_{r}}}.
\end{equation}
Here $U$ and $T$ respectively are the internal energy and the temperature of the system and $R$ is the gas constant. Further, one restricts to the electronic excitations of the atoms. In the present context however we shall consider, in analogy, the quark excitations of the quarkons and deal with the transfer of one unit of color (anticolor) charge at a time from parent to daughter quarkon or the vice-versa.

\par	
	It may be noted that for a simple atomic process for an element $M$, the ionization and capture described by $M,M$ $\rightleftharpoons$ $M^{+} + e^{-}$, ($M^{+}$ is the ion and $e^{-}$ is the electron), equation (1) is written as[12]

\begin{equation}
 ln \frac{P_{M^{+}}}{P_{M}} P_{e^{-}} = -\frac{U}{RT} + \frac{5}{2} ln T + \kappa + ln (g_{e} \frac{Z_{e}(M^{+})}{Z_{e}(M)}),
\end{equation} 
where $g_{e}$ is the electron spin multiplicity,i.e.,$2$ and $Z_{e}$ denotes the partition function of the atom in the argument , and the constant $\kappa$ is given by $\kappa = ln((2 \pi m)^{3/2} k^{5/2}/h^{3})$. Under the assumption that one initially starts with the element M and then goes on heating the system in a confined space to a temperature $T$, the fraction $x$ of the ionized atoms, also linked with the partial pressures $P_{M}$, $P_{M^{+}}$,$P_{e^{-}}$ through $P_{e}=P_{M^{+}}= nxkT$, $P_{M}= n(1-x)kT$, can be computed from (3) as

\begin{equation}
 ln \frac{x^2}{(1-x^2)}P = -\frac{U}{RT} + \frac{5}{2} ln T - 14.875, 
\end{equation}
where $P = P_{e^{-}}+P_{M^{+}}+P_{M}$ is the total pressure and the contribution of the last term in (3) is ignored. Further, formula (3) can also be written for the ionization of the atom at any stage of ionization, namely from $M^{r}$ to $M^{r+1}$ (where $M^{r}$ denotes the atom which has already lost $r$-electrons), as

\begin{equation}
 ln \frac{n_{r+1}}{n_{r}} P_{e} = -\frac{U_{r}}{RT} + \frac{5}{2} ln T + \kappa + ln (2 \frac{Z_{e}(M^{r+1})}{Z_{e}(M^{r})}).
\end{equation} 
Here $U_{r}$ is the heat of ionization from $M^{r}$ to $M^{r+1}$ and the electronic partition function of $M^{r}$, $Z_{e}(M^{r})$, for the $r$-times ionized atom is given by

\begin{equation}
 Z_{e}(M^{r}) = g_{r} + \sum_{s}g_{rs} e^{-\chi_{s}/T},
\end{equation}
where $g_{r}$ and $g_{rs}$ are being the weights corresponding to the ground and $s-$th excited state of $M^{r}$; $\chi_{s}$ is the excitation energy of state $s$, and the temperature T is now onward expressed in the units of $k$.
\par
	For the sake of ready use we give below another version of formula (1), namely[13] for the reaction $\sum_{i} \nu_{i} A_{i} = 0$, one writes (1) in the form

\begin{equation}
 K(P,T) = P^{-\sum_{i} \nu_{i}}.e^{-\sum_{i} \nu_{i}\chi_{i}/T}
\end{equation}
which, for a paricular reaction $A^{+}+e^{-}-A^{0} = 0$ with $\nu_{A^{+}} = 1$, $\nu_{e^{-}} = 1$, $\nu_{A^{0}}= -1$, reduces to the form 

\begin{equation}
 K(P,T) = \frac{1}{P}.\frac{g_{e}g_{+}}{g_{A}} (\frac{2\pi m}{h^2})^{3/2}.T^{5/2} e^{-\chi_{s}/T},
\end{equation}
or after using $P = (N/V)T$, one writes (8) as

\begin{equation}
 K(P,T) = \frac{V}{N}.\frac{g_{e}g_{+}}{g_{A}} (\frac{2\pi m}{h^2})^{3/2}.T^{3/2} e^{-\chi_{s}/T}.
\end{equation}
In this case, if one defines the degree of ionization (ionic fraction) $\alpha = N_{+}/N_{0}$, then eq.(9) can be recast (see, Rumer and Ryvkin, Ref.(13)) as $\frac{\alpha^2}{(1-\alpha)} = 2 \phi(T,V)$, leading to

\begin{equation}
 \alpha(T,V) = -\phi(T,V) + (\phi(T,V)^2 + 2\phi(T,V))^{1/2}
\end{equation}
where
\begin{equation}
 \phi(T,V) = (T/T_{0})^{3/2}.e^{-\chi_{s}/T}, with T_{0}= (\frac{h^2}{2\pi m})(2\frac{g_{A}}{g_{e}g_{+}}\frac{N_0}{V})^{2/3},
\end{equation}
is a positive, dimensionless function. Note the difference between the definitions of fractions $\alpha$ in eq.(10) and $x$ of eq.(4). We shall however investigate $\alpha$ as a function of $T$.

\par
	We shall restrict here only to two-body final states and use $g_{e}=2$ for quarks and $g_{e}=1$ for scalar diquarks. Note that the formula (10) is more convenient if the ionized gas is confined to a fixed volume as is the case with fire-ball in RHIC experiments. Further,we shall use the version of SIF as described in eqs. (8)-(11) for quarkons and diquarkons, which in certain situations will correspond to pentaquark baryons and tetraquark mesons. The picture considered here is as follows:
\par	
	Whether it is deconfined phase of quarks and gluons that existed at the time of early Universe or the same is produced in RHIC experiments in a localized region (fire-ball), the system cools down and evolves in space and time thereafter. As a result, the hadronization takes place after the system passes through (may be for a short while) a diquark-gluon phase or a mixed quark-diquark-gluon phase. Alternatively, one can think of (at least in the case of RHIC experiments in the forward process) the breaking of the normal hadrons at such high energies into their constituent quarks and thereby simultaneously forming a metastable state of quark and/or diquark clusters. We assume that quarks or diquarks in these clusters are bound by Coulomb-like color forces, as the corresponding potential is found to work well for quark constituents of the nucleon[14] and also of pentaquark baryons[8]. Note that pentaquark baryons or tetraquark mesons could be one of the possibilities of these quarkons or diquarkons present initially. In the next stage , these clusters will decay and give rise to a mixture of quarks and diquarks and finally to quarks only. With regard to this picture, we make the following simplifying but plausible assumptions: (i) Role of gluons in this model will appear only through the transfer of color charge; (ii) The transfer of one quark will carry only one unit of color charge ( as is the case with the Coulombic charge of electron) and that of a diquark will carry two units of color charge; (iii) No account will be made of fractional Coulomb-charge on a quark or diquark (or on the quarkon or diquarkon for that matter); (iv) We also ignore an account of nature of color on a particular quark and also that of its flavor for the time being. (v) While neglecting other production or recombination processes, we shall concentrate here only on the decay of quarkons and diquarkons into lighter quarkon channels -perhaps more justified for the formation of QGP. Some of these assumptions conform to the spirit of color-quark-chemistry studied earlier by Chan Hong-Mo et al.[11].

\section{Quarkons and Diquarkons in the Fire-ball and their Decay}
\subsection{The processes involving pentaquark baryons and tetraquark mesons}
The quarkons and diquarkons, once formed in the fire-ball in RHIC experiments, subsequently decay into quarks and diquarks and in due course contribute to the deconfined state of these objects and that too in a confined volume. While the studies can be easily extended to a general case , for simplicity we concentrate here only on the decay of pentaquarkons and tri-diquarkons, denoted respectively by $Q_{(0)}^{(5)}$ and $D_{(0)}^{(3)}$. Here superscript denotes the number of quarks(diquarks) present initially in the cluster (termed as 'parent') and subscript represents the number of quarks (diquarks) left over after the decay (termed as 'daughter'). In other words, the quarks (or for that matter diquarks) released finally from the cluster due to the increase in temperature of the fire-ball, will be able to keep a track of their parentage. The possible processes associated with the decay of these systems can be listed as follows:

\begin{equation}
 Q_{(0)}^{(5)} \rightarrow Q_{(4)}^{(5)} +Q_{(1)}^{(5)},
\end{equation}   
\begin{equation}
               \rightarrow Q_{(3)}^{(5)} +Q_{(2)}^{(5)}\rightarrow Q_{(3)}^{(5)}+2Q_{(1)}^{(5)},
\end{equation} 

\begin{equation}
              \rightarrow Q_{(2)}^{(5)} +Q_{(3)}^{(5)}\rightarrow Q_{(2)}^{(5)}+3Q_{(1)}^{(5)},
\end{equation} 
\begin{equation}
              \rightarrow 2Q_{(2)}^{(5)} +Q_{(1)}^{(5)}\rightarrow 5Q_{(1)}^{(5)},
\end{equation} 
\begin{equation}
 Q_{(0)}^{(4)} \rightarrow Q_{(3)}^{(4)} +Q_{(1)}^{(4)},
\end{equation} 
\begin{equation}
               \rightarrow Q_{(2)}^{(4)} + 2Q_{(1)}^{(4)},
\end{equation} 
\begin{equation}
               \rightarrow 2 Q_{(2)}^{(4)}\rightarrow 4Q_{(1)}^{(4)},
\end{equation} 
\begin{equation}
 Q_{(0)}^{(3)} \rightarrow Q_{(2)}^{(3)} +Q_{(1)}^{(3)},
\end{equation} 
\begin{equation}
               \rightarrow 3 Q_{(1)}^{(3)},
\end{equation} 
\begin{equation}
 Q_{(0)}^{(2)} \rightarrow Q_{(1)}^{(2)} +Q_{(1)}^{(2)},
\end{equation} 
\begin{equation}
 D_{(0)}^{(3)} \rightarrow D_{(2)}^{(3)} +D_{(1)}^{(3)},
\end{equation} 
\begin{equation}
               \rightarrow  3 D_{(1)}^{(3)},
\end{equation} 
\begin{equation}
 D_{(0)}^{(2)} \rightarrow 2 D_{(1)}^{(2)},
\end{equation} 
\begin{equation}
 D_{(0)}^{(1)} \equiv Q_{(0)}^{(2)} \rightarrow 2 Q_{(1)}^{(2)},
\end{equation} 

\par
	While the processes (22)-(24) will contribute to the diquark or quark-diquark plasma, the processes (12) to (21) and (25) will be responsible exclusively for quark plasma. In fact the role of pentaquark baryons or tetraquark mesons, if at all manifests in the formation of QGP, it is reasonable to assume that the same will appear through the decays (color ionization) of pentaquarkons $Q_{(0)}^{(5)}$, tetraquarkons $Q_{(0)}^{(4)}$, $Q_{(4)}^{(5)}$, $D_{(0)}^{(2)}$ including that of other parent clusters; whereas the decay of triquarkons $Q_{(3)}^{(4)}$, $Q_{(3)}^{(5)}$, diquarkons $D_{(1)}^{(2)}$, $Q_{(2)}^{(4)}$, $Q_{(2)}^{(5)}$, and monoquarkons $Q_{(1)}^{(4)}$,
$Q_{(1)}^{(5)}$ will contribute to quark and diquark plasma. We compute here the color ionic fraction (CIF) for some representative cases using SIF for the colored ions and study the same as a function of temperature at various quark number densities. 

\subsection{Calculation of color ionic fraction for various processes}
When computing the CIF using (10) and (11), we need some ingredients about the quark composites. The same we use from our earlier works[8,14] in which a quark diquark (QDQ) model for the nucleon[14] and a quark double diquark (QDDQ) model for the pentaquark baryons[8] are proposed. With regard to the geometry of the fire-ball (though it depends on the colliding ions) in a typical RHIC experiment, the results are taken from Karsch and Petronzio [15] and others[16] derived in the context of $J/\psi$ suppression.

\par
	In particular, we concentrate on some typical processes (12), (13), (18), (19) and (21). In fact these are the cases for which some parameters about the concerned quarkons/diquarkons are already known[8,14]. Note that in the same spirit we consider here Coulomb-like color forces among the quarks in a quarkon and among the diquarks in a diquarkon and use the hydrogenic model for deriving the various parameters from the experimental results. In this way the ionization potential,$\chi_{s}$, in (11) for various clusters are computed. Thus, we use the following ingredients:\\

$\bf{QDQ Model[14]:}$

\begin{eqnarray}
 m_{q}= 513.0 MeV; m_{D}= 681.3 MeV; b_{0c}=0.47 fm; b_{0v}= 0.57 fm;  \nonumber\\
 |\epsilon_{1}^{c}|=349.0 MeV;  |\epsilon_{1}^{v}|= 215.8 MeV; \beta_{c}= 1.64; \beta_{v}= 1.23. 
\end{eqnarray} 

$\bf{QDDQ Model[8]:}$

\begin{eqnarray}
B_{0c}= 0.8 fm; B_{0v}= 0.48 fm; \delta_{c}=0.723 ; \delta_{v}=1.13 ; \nonumber\\
 |E_{1}^{c}|= 90.4 MeV;  |E_{1}^{v}|= 235.4 MeV .
\end{eqnarray} 

\par
	For the tetraquark (a state of diquark-antidiquark ($D\bar{D}$) bound system) case, we assume the following relationships for couplings:

 \[\frac{\beta_{q\bar{q}}}{\beta_{qq}} = \frac{\delta_{D\bar{D}}}{\delta_{DD}},\]

and derive $\delta_{D\bar{D}}$ using the models described in Refs.(8) and (14). This leads to $\delta_{D\bar{D}}= 1.06$ for light quarks and $\delta_{D\bar{D}}= 0.56$ for the heavy (charm) quarks and accordingly for the tetraquark meson we find[8] $|E_{1}^{D\bar{D}}|= 191.4 MeV$, for the light $D\bar{D}$-case, and $|E_{1}^{D\bar{D}}|= 54.2 MeV$,for the heavy (charm)$D\bar{D}$-case. Next we calculate the CIF, $\alpha$, from (10) for the following four cases:\\

$\bf{Case Ia}$: Decay of the pentaquarkon (cf. eq.(12), valence):$Q_{(0)}^{(5)}\rightarrow Q_{(4)}^{(5)}+Q_{(1)}^{(5)}$\\
\par
	In this case the quark is released from the valence in the QDDQ model of the pentaquark baryons[8] and the process will contribute to the quark gas (cf. Fig. 1a).\\

$\bf{Case Ib}$: Decay of the pentaquarkon (cf. eq.(13), core):$Q_{(0)}^{(5)}\rightarrow Q_{(3)}^{(5)} +Q_{(2)}^{(5)}$\\
\par
	This is the case in which one of the diquarks from the core in the QDDQ model[8] is released. This process will contribute to the diquark gas (cf. Fig. 1b).\\

$\bf{Case IIa}$: Decay of the triquarkon (cf. eq.(19), valence):$Q_{(0)}^{(3)}\rightarrow Q_{(2)}^{(3)}+Q_{(1)}^{(3)}$\\
\par

	In this case the triquarkon is ionized and will contribute to both quark and diquark gases or to their mixture. The valence in the QDQ model of nucleon[14] is released (cf. Fig. 2a).\\

$\bf{Case IIb}$: Decay of the triquarkon (cf. eq.(19), core):$Q_{(0)}^{(3)}\rightarrow Q_{(2)}^{(3)} +Q_{(1)}^{(3)}$\\
\par
	This is the case in which the quark from the diquark core of the nucleon[14] is released
and will contribute to both quark and diquark gases or to their mixture (cf. Fig. 2b).\\

$\bf{Case III}$: Decay of tetraquarkon (cf. eq.(18):$Q_{(0)}^{(4)}\rightarrow Q_{(2)}^{(4)}+Q_{(2)}^{(4)}$\\
\par
	In this case, tetraquarkon breaks up into two diquarks and will exclusively contribute to diquark gas and the model of Ref.(14) is used (cf. Fig. 3).\\

$\bf{Case IV}$:Decay of diquarkon (cf. eq.(21):$Q_{(0)}^{(2)}\rightarrow Q_{(1)}^{(2)}+Q_{(1)}^{(2)}$\\
\par
	This process will exclusively contribute to quark gas and the parameters are derived by assuming the diquark to be the same as in the core of the nucleon in the QDQ model[14].\\

\par
	Corresponding to these processes the CIF is computed from eq. (10). The values of various parameters used in the calculations are those given in eqs. (26) and (27) and in Table 1 for a sample case of $O^{16}$-$O^{16}$ collision. For this case, an order of magnitude estimate for the size of the fire-ball and of $(N_{0}/V)$ are carried out using the parameters given in Refs.(15) and (16). The calculated reduced mass, $m$, of the two-body final state and the spin multiplicities $g_{e}$, $g_{+}$, $g_{A}$ along with the calculated ionization energy $\chi_{s}$,are listed in Table 1 for different cases. Further, for the above six cases the calculated results for the CIF,$\alpha$, as a function of temperature $T$ are shown in Figs.1-4  corresponding to the three typical values of the particle number density $(N_{0}/V)$, namely $10$x$10^{6}$ (continuous curve),$50$x$10^{6}$ (dashed curve) and $90$x$10^{6}$ $MeV^3$ (dotted curve). While $T=200 MeV$ is believed to be the temperature for the hadronization, the $\alpha$ is shown in these figures in the range $T=50 MeV$ to $350 MeV$. Some crucial difference around this temperature can be seen from these figures for differnt cases and for different particle number densities.\\

\begin{table}
\caption{Values of various quantities used in the calculations of color-ionic fraction, $\alpha$, from eqns.(10) and (11).}

\vspace{0.1in}
\begin{tabular}{|p{2.0cm}|p{1cm}|p{1cm}|p{1cm}|p{1.5cm}|p{1.5cm}|p{1.5cm}|}
\hline
$process$&$g_{e}$&$g_{+}$&$g_{A}$&$m(MeV)$&$\chi_{s}(MeV)$&$T_{0}(MeV)$\\
\hline
Case Ia&2&1&2&400.0&235.4&3026.0\\
Case Ib&1&2&2&460.0&90.4&1664.0\\
Case IIa&2&1&2&288.3&215.8&4225.0\\
Case IIb&2&1&2&333.3&349.0&3656.0\\
Case III&1&1&1&340.5&191.4&2257.0\\
Case IV&2&2&1&250.5&349.0&1935.0\\
\hline
\end{tabular}
\end{table}

\section{A Connection between Color-Ionization and Nucleation Mechanism}
Nucleation or precipitation mechanism in a medium has been the subject of study in various field and in different contexts for many decades now[17-19]. The nucleation rate in the context of QCD and QGP has been investigated by Csernai and Kapusta[20] among others[21]. In this case one computes the probability that a bubble or a droplet of $A$-phase appears in a system initially in the $B$-phase near the critical temperature. Again for the case like early universe or RHIC studies homogeneous nucleation theory is found more convenient. In fact a droplet of critical size is metastable, it is balanced between evaporation and accretion. Somewhat similar is the picture in the present case in which a quarkon (or a diquarkon) of critical size ( in the sense that it becomes a normal hadron for a certain number of quarks(diquarks) and a part of the medium just for another number) plays the role of a droplet or bubble. The decay of quarkon or diquarkon in this case, can be considered as the process of evaporation of a droplet. Further, for the study of  nucleation rate both classical theory[17,18], based on thermodynamics, and the modern theory of Langer[19], based on statistical methods have been used. The concept of nucleation from one vacuum to another has also been extended to the domain of quantum field theory by several authors[22].

\par
	The interesting part in all these studies in rather disconnected fields is that the derived expression for the nucleation rate, in general, has a common feature in terms of physical as well as mathematical contents. While the probability per unit time per unit volume to nucleate the dense 'liquid' phase from a dilute 'gas' phase is broadly expressible as

\begin{equation}
     I = I_{0}.e^{-W_{c}/T},
\end{equation}
the prefactor $I_{0}$ (having dimensions of $T^4$) here in general has a break-up into statistical and dynamical parts in all the theories of nucleation. For example, in the Langer's theory[19], one writes the nucleation rate as

\begin{equation}
     I = \frac{\kappa}{2\pi} \Omega_{0}.e^{-\Delta F/T},
\end{equation}
where $\Delta F$ is the change in the free energy of the system due to the formation of critical droplet, and $\Omega_{0}$ and $\kappa$ are the statistical and dynamical prefactors which respectively are the measures of the available phase space volume and the exponential growth of the critical droplet. For the theories of early Universe while $I_{0}$ in the literature( see, for example,[21]) is given by $I_{0}=(W_{c}/2\pi T)^{3/2} T^4$, Csernai and Kapusta within the frame work of a course-grained effective field theory approximation, obtain an expression for $I_{0}$ in the QGP context as[20,21]

\begin{equation}
    I_{0} = \frac{16}{3\pi} (\frac{\sigma}{3T})^{3/2} \frac{\sigma\eta_{q} R_{c}}{(\xi_{q})^4 (\Delta w)^2},
\end{equation}
where $\eta_{q}= 14.4 T^3$, is the shear viscosity in the plasma phase;$\xi_{q}$ is the correlation length in this phase; $\sigma$ is the surface free energy; $R_{c}$ is the radius of the critical-sized bubble, and $\Delta w$ is the difference in the enthalpy densities of the two phases.

\par
	It is worthwhile to compare the formula (29) (or (28)) with $I_{0}$ given by (30) with the
expression (9) defining the fraction of partial pressures for the products and the reactants, particularly for the single color-ionization. First of all note that beside the factor $(V/N)$ on the right hand side of (9), the factor $(g_{e}.g_{+}/g_{A}$ has a statistical origin and corresponds to $\Omega_{0}$ in (29). The remaining factor $(2\pi mT/h^2)^{3/2}$ is analoguous to $\kappa$ in (29) as it has roots in the dynamics. Secondly, a little simplification of the prefactor $I_{0}$ in (30) yields

\begin{equation}
    I_{0} = (T/\bar{T_{0}})^{3/2}, with  \bar{T_{0}}=(\frac{\pi \xi_{q}^4 (\Delta w)^2}{230.4 R_{c}})^{2/3} (\frac{3}{\sigma})^{5/3}.
\end{equation}   
On the other hand, the prefactor, $K_{0}$, in (9) can also be expressed as
\begin{equation}
    K_{0} = (T/\tilde{T_{0}})^{3/2}, 
\end{equation}
where $\tilde{T_{0}}$ is given by
\begin{equation}
 \tilde{T_{0}} = (\frac{N_{0} g_{A}}{V g_{e}g_{+}})^{2/3} (\frac{h^2}{2\pi m}),
\end{equation}  
and $K$ can still be shown to have dimensions of $T^4$ like $I$ in (28), at least for the case of single color-ionization where $K$ turns out to be a measure of pressure. Further it may be of interest to compare the magnitudes of the ionization energy $\chi_{s}$ (cf. Table 1) and that of $\Delta F$ in (29). In fact they turn out to be of the same order (cf. Ref. (21)) in certain models. 
\par
	Thus, seemingly different mechanisms of single color-ionization and bubble nucleation have the same mathemathical content at their computational level.

\section{Discussion and Conclusions}
From the point of view of investigating the role of newly discovered pentaquark baryons and tetraquark mesons in the formation of QGP, we have made a modest attempt, perhaps for the first time, to demonstrate the viability of SIF for the case of colored ion systems. While the method is quite general for the study of multiply ionized systems, the case of single (color-) ionization is investigated in detail. In spite of several assumtions made to simplify the computation of CIF, this latter quantity as a function of temperature yields the behaviour as expected. Interestingly,
for this single-ionization case a connection of the prsent approach with the well studied problem of bubble nucleation in the context of QGP is demonstrated. There appears to be a one-to-one correspondence between the working formulae in the two approaches, in spite of different physical inputs in the two cases. To be more specific, our findings can be summarized as follows:\\

(i) The fact that CIF, $\alpha$, in all cases (cf. Sect.3.2) at different particle number densities approaches to unity in the large-$T$ limit, implies a complete dissociation of quarks or diquarks from the corresponding clusters, i.e., the formation of a noninteracting quark or diquark gas.\\ 

(ii) Note from Figs. 1a and 1b that at a given temperature, the contribution to diquark gas from the diquarks in the core of a pentaquarkon is more pronounced compare to the one obtained from its valence quark within the frame work of the QDDQ model. This conforms to the fact that the valence of antiquark in the pentaquark baryon is more tightly bound compare to the double diquark core -an important outcome of the QDDQ model[8].\\

(iii) In case of triquarkons, the core and valence (cf. Ref.(14)) contributions to both diquark and quark gases are comparable as far as the variation of $\alpha$ with $T$ is concerned. However, the difference manifests in the low-T limit (cf. Figs. 2a and 2b).\\

(iv) As expected, the diquarks of tetraquarkons contribute somewhat more to the diquark gas(cf. Fig. 3) compared to the diquarks of triquarkons (cf. point (iii) above).\\

\par
	As the colored-ion system (such as the fire-ball in RHIC experiments) evolves in space and time, the temperature (cf. Refs.(15) and (16)) and hence the CIF (cf. eq.(10)) also varies inside the fire-ball with space and time. This kind of space and/or time dependence of $\alpha$ can easily be investigated in the present frame work of SIF and that too for different stages of multiple ionization of quarkons and diquakons. However, for siplicity we have taken $T$ as uniform all through out the volume of the fire-ball.

\par
	Note that it is only the single (color-)ionization in the generalized version (1) of SIF which corresponds to the bubble nucleation mechanism in QGP (cf. Refs. (20) and (21)). This shows the richness of SIF approach over the nucleation one as far as the hadronic-bubble formation in QGP is concerned . Moreover, there is enough scope to investigate further higher-order phase transitions in QGP (if at all they exist) in the present approach by way of studying the two- or higher-stages of ionization of quarkons and/or diquarkons. Such studies are in progress.

\begin{acknowledgements}
Initially, this work was presented in the ICPAQGP-2005, held at Kolkata during Feb. 8-12, 2005. The author wishes to acknowledge the comments of various delegates to the Conference. Final touch to this work was given during author's visit to IUCAA, Pune. He thanks Professors N. K. Dadhich and A.K.Kembhavi for the facilities. Thanks are also due to Dr. D. Parashar for a critical reading of the manuscript.
\end{acknowledgements}

\newpage
{\bf \centerline {REFERENCES}}
\vspace{1cm}

[1]. Proc. of fourth Int. Conf. on {\sf Physics and Astrophysics of Quark-Gluon Plasma} ed. by B. Sinha, D. Srivastava and Y.P. Viyogi (Ind. Aca. Sc. Bangalore, 2000), Sp. Issue of Pramana- J. Phys. {\bf 60} (2003) p. 577-1135.\\

[2]. Proc. of third Int. Conf. on {\sf Physics and Astrophysics of Quark-Gluon Plasma} ed. by B. Sinha, D. Srivastava and Y.P. Viyogi (Narosa Publishing House, New Delhi, 1998)p.1-644.\\

[3]. See, for example, J. Cleyman and H. Satz, Z. Phys. {\bf C57} (1993) 135; C.S. Gao and T. Wu, J. Phys. G: Nucl. and Part. Phys. {\bf 27} (2001)459.\\

[4]. J. F. Donoghue and K. S. Sateesh, Phys. Rev. {\bf D38} (1989) 360.\\

[5]. D. Kaster and J. Traschen, Phys. Rev. {\bf D44} (1991) 3791.\\

[6]. S. K. Karn, R. S. Kaushal and Y. K. Mathur, Eur. Phys. Jour. {\bf C14} (2002) 487; S. K. Karn, R. S. Kaushal and Y.K. Mathur, Z. Phys. {\bf C72} (1996) 297, and the references therein.\\

[7]. T. Nakano et al, LEPS Collaboration, Phys. Rev. Lett. {\bf 91} (2003) 012002; A. Aktas et al.,H1 Collaboration, hep-ex/040317; C. Alt et al., NA49 Collaboration, Phys. Rev. Lett. {\bf 92} (2004) 042003; V. Kubarovsky et al, CLAS Collaboration, Phy. Rev. Lett. {\bf 92} (2004) 032001.\\

[8]. R. S. Kaushal, D. Parashar,and A.K. Sisodiya, Phys. Lett. {\bf B600} (2004) 215.\\

[9]. Y. Oh, H. Kim and S. H. Lee, Phys. Rev. {\bf D69} (2004) 014009; Q. Zhao, Phys. Rev. {\bf D69} (2004) 053009; M. Karliner and H. J. Lipkin, Phys. Lett. {\bf B575} (2003) 249; M. Karliner and H. J. Lipkin, hep-ph/0307243; K. Maltman, Phys. Lett. {\bf B604} (2004) 175.\\

[10]. L.-W. Chen,V. Greco, C. M. Ko, S.H. Lee and W. Lin, Phys. Lett. {\bf B601} (2004) 34.\\

[11]. Chan Hong-Mo and H. Hogaasen, Nucl. Phys. {\bf B136} (1978) 401; Phys. Lett. {\bf B72} (1977) 121, Chan Hong-Mo et al, Phys. Lett. {\bf B76} (1978) 634.\\

[12]. M. N. Saha and B. N. Srivastava,{\sf A Treatise on Heat} (The Indian Press Pvt. Ltd., Allahabad, reprinted ed. 1972 and first ed. 1931); Also see , M.N. Saha and Rai, Proc. Nat.Aca. Sc. {\bf 4}, 319; B.N. Srivastava, Proc. Roy. Soc. Lond. {\bf 175} (1940) 26.\\

[13]. See, for example, Y. B. Rumer and M. S. Ryvkin,{\sf Thermodynamics,Statistical Physics and Kinetics} (Mir Publishers, Moscow, 1980) pages 254,158; V. B. Bhatia, {\sf Textbook of Astronomy and Astrophysics with Elements of Cosmology} (Narosa Publishing House, New Delhi, 2001) p. 37.\\

[14]. R. S. Kaushal and D. S. Kulshreshtha, Ann. Phys.(NY) {\bf 108} (1977)198; R. S. Kaushal, Phys. Lett. {\bf B57}(1975) 354; {\bf B60} (1975) 181.\\

[15]. F. Karsch and R. Petronzio, Phys. Lett. {\bf B212} (1988) 255; Z. Phys. {\bf C37} (1988) 627.\\

[16]. R.S. Kaushal and S. K. Karn, Pramana-J. Phys. {\bf 44} (1995) 167.\\

[17]. R. Becker and W. Doring, Ann. Physik (Leipzig) {\bf 24} (1935) 719.\\

[18]. J. E. McDonald, Am. Jour. Phys. {\bf 30} (1962) 870; {\bf 31}(1962) 31.\\

[19]. J. S. Langer, Ann. Phys. (NY) {\bf 54} (1969) 258.\\

[20]. L. P. Csernai and J. I. Kapusta, Phys. Rev. Lett. {\bf 69} (1992) 737; Phys. Rev. {\bf D46} (1992) 1379.\\

[21]. See, for example, D. Chandra and A. Goyal, Int. Jour. Mod. Phys. {\bf A19} (2004) 5221.\\

[22]. C. G. Callan and S. Coleman, Phys. Rev. {\bf D16} (1977) 1762; I. Affleck, Phys. Rev. Lett. {\bf 46} (1981) 388; A. D. Linde, Nucl. Phys {\bf B216} (1983) 421.\\ 


\newpage

\begin{figure}
\figurename{1a: Contribution to quark gas from pentaquarkons (valence): The results are shown for three values of paricle number densities $(N_{0}/V)$, namely $10$x$10^6$, $50$x$10^6$ and $90$x$10^6 MeV^3$, respectively by continuous, dashed and dotted curves.}
\begin{center}
\includegraphics[angle=270, width=5in]{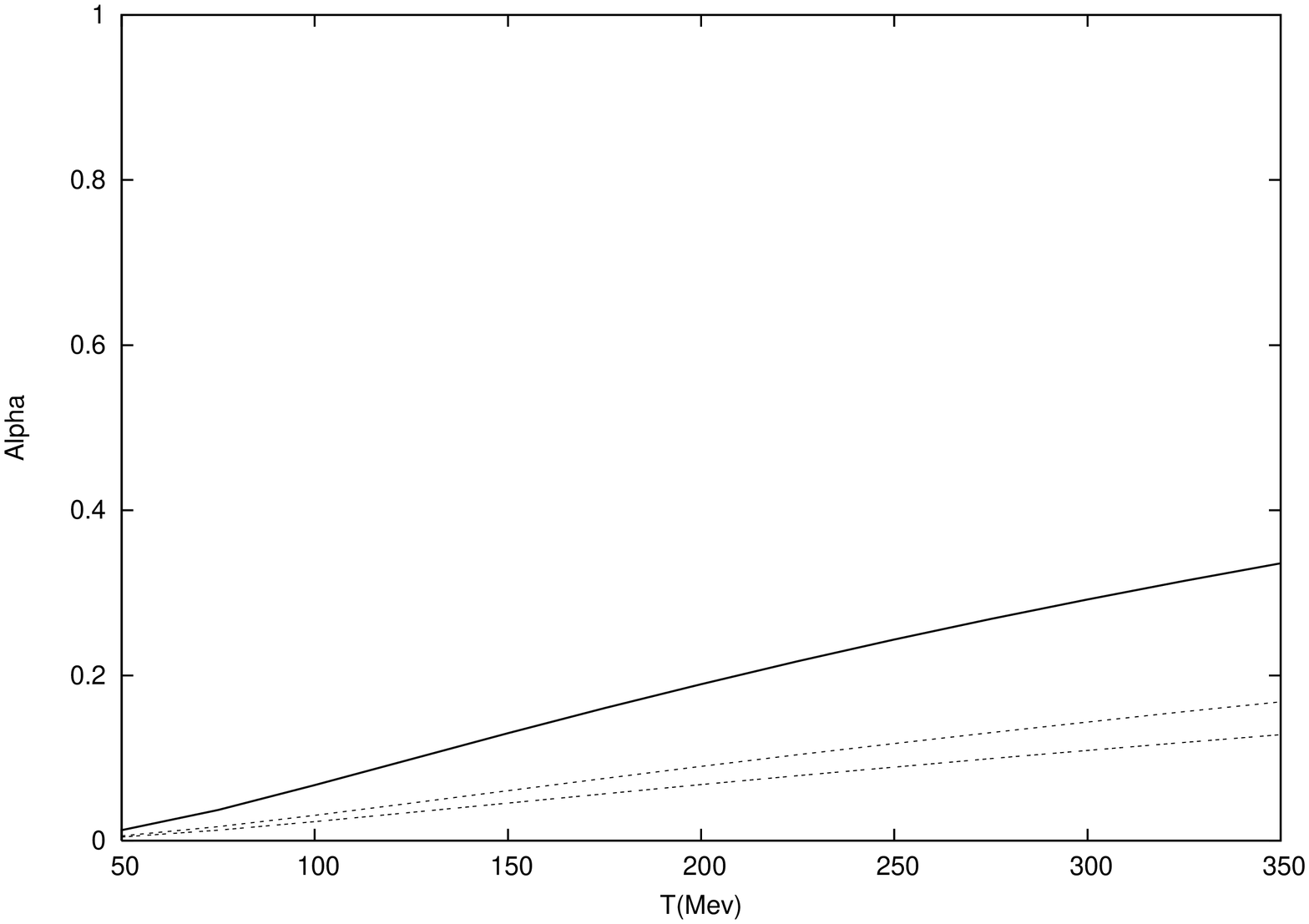}
\end{center}
\end{figure}

\begin{figure}
\figurename{1b: Contribution to diquark gas from pentaquarkons (core). The description of curves is the same as in Fig. 1a.}
\begin{center}
\includegraphics[angle=270, width=5in]{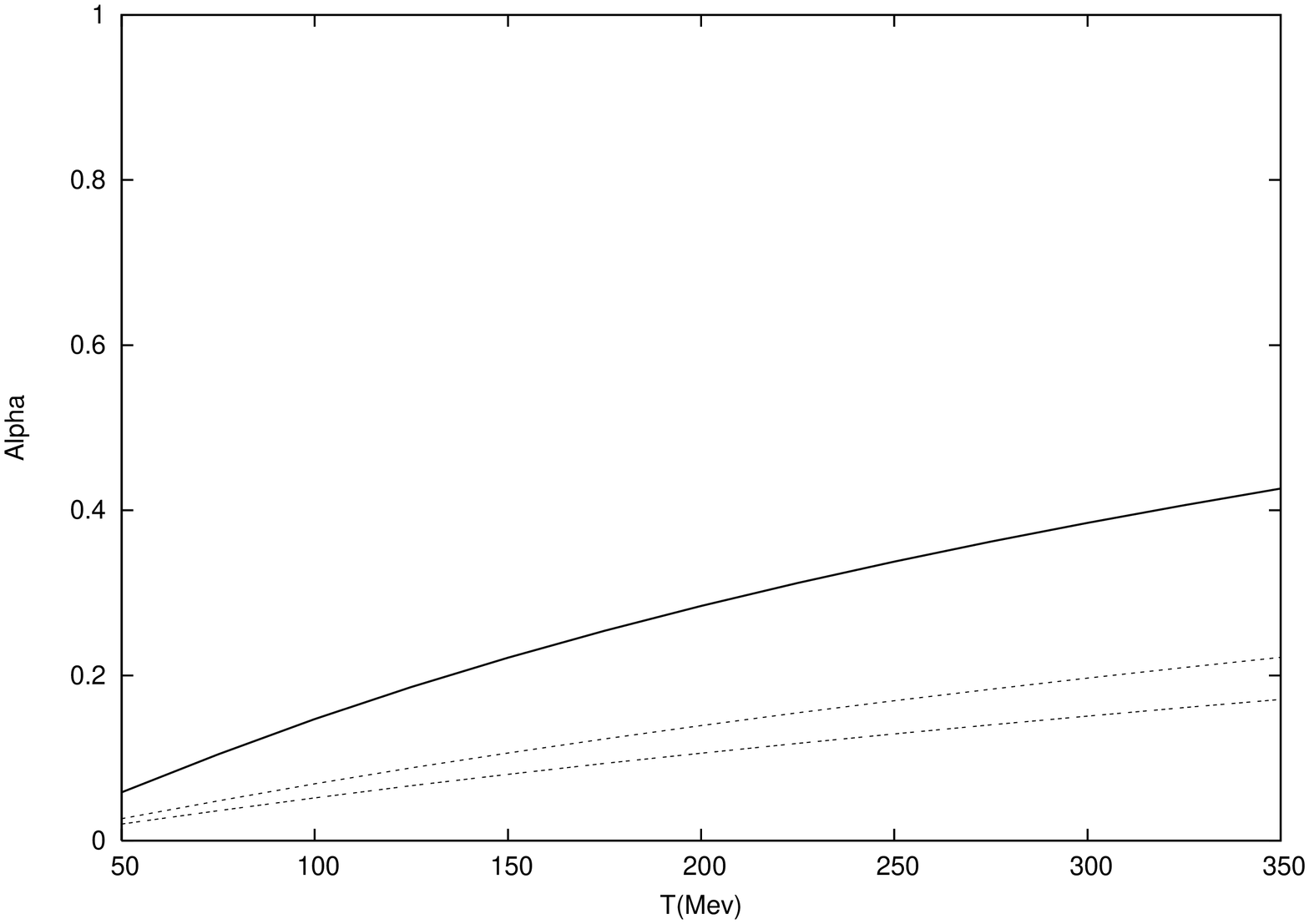}
\end{center}
\end{figure}

\begin{figure}
\figurename{2a: Contribution to quark and diquark gas from triquarkons (valence). The description of curves is the same as in Fig. 1a.}
\begin{center}
\includegraphics[angle=270, width=5in]{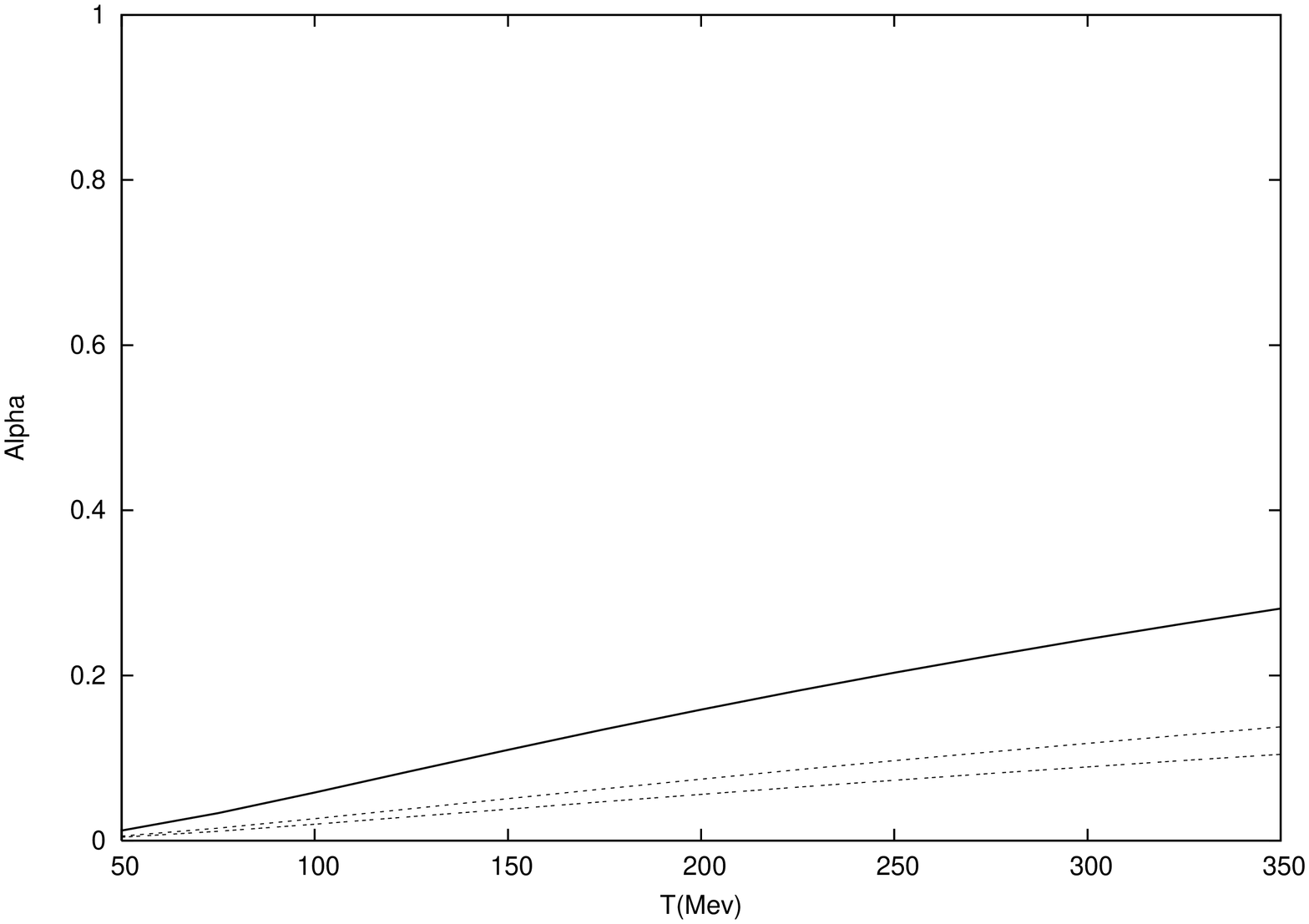}
\end{center}
\end{figure}

\begin{figure}
\figurename{2b: Contribution to quark and diquark gas from triquarkons (core). The description of curves is the same as in Fig. 1a.}
\begin{center}
\includegraphics[angle=270, width=5in]{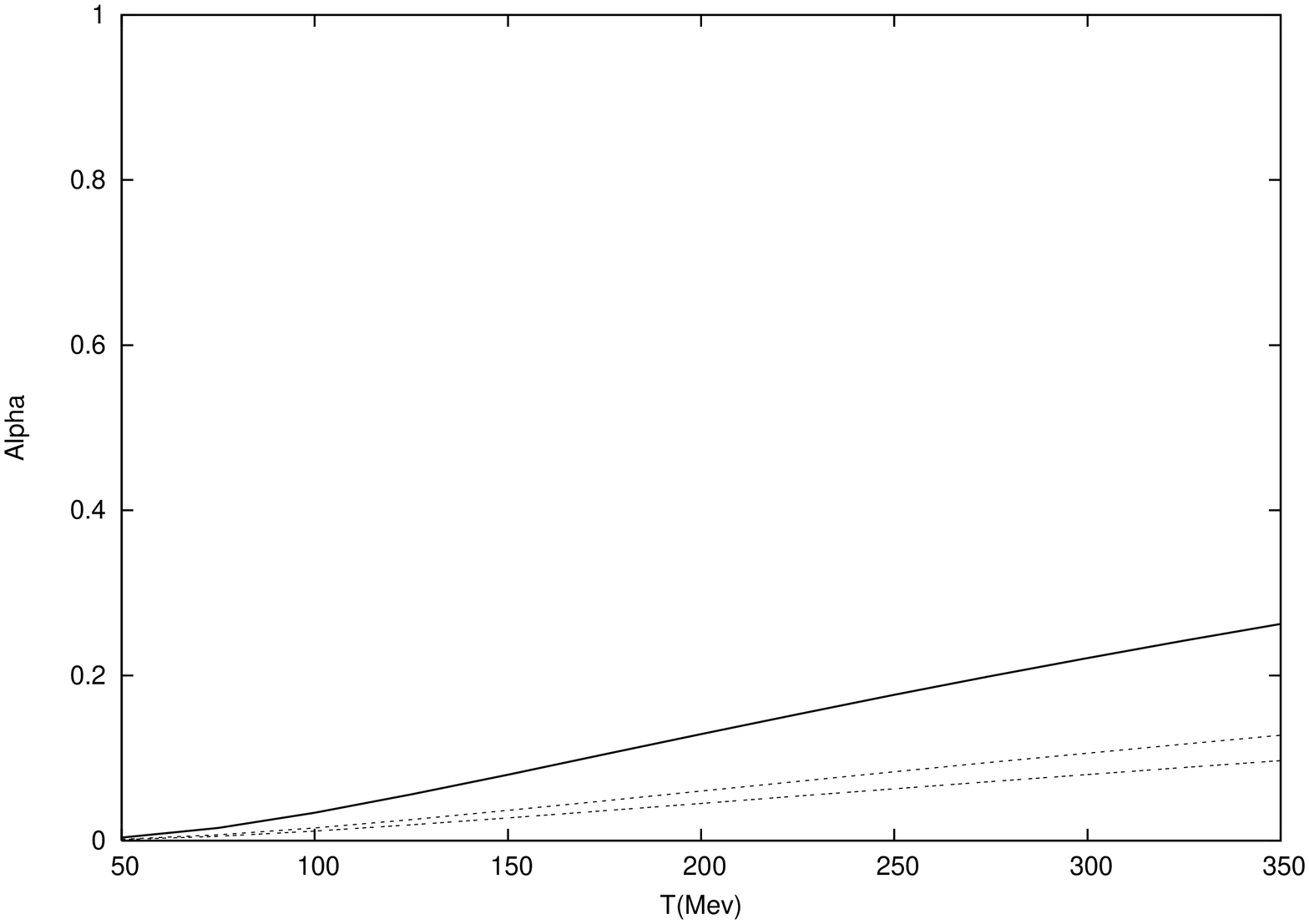}
\end{center}
\end{figure}

\begin{figure}
\figurename{3: Contribution to diquark gas from tetraquarkons. The description of curves is the same as in Fig. 1a.}
\begin{center}
\includegraphics[angle=270, width=5in]{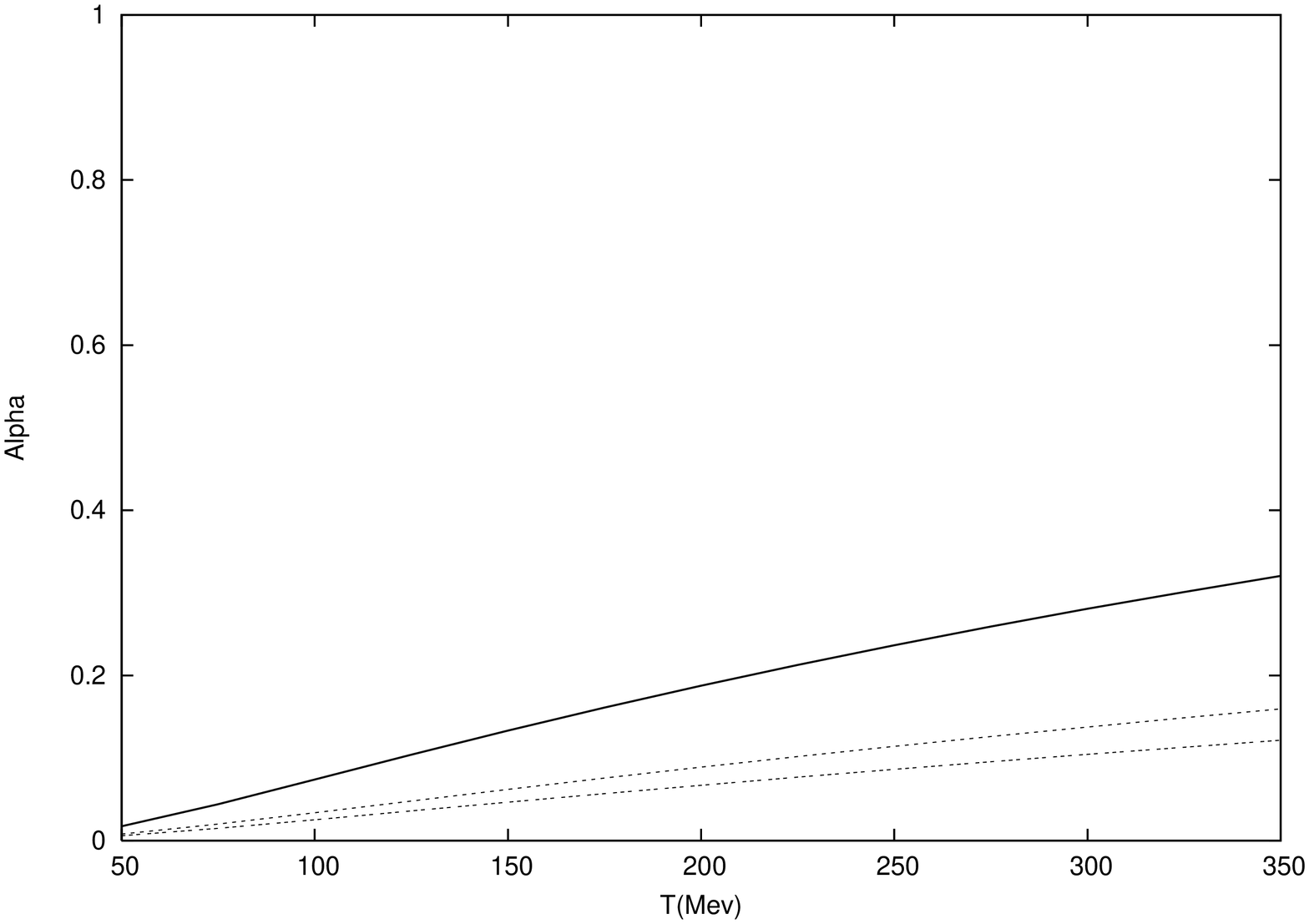}
\end{center}
\end{figure}

\begin{figure}
\figurename{4: Contribution to quark gas from single diquarkons (core-like). The description of curves is the same as in Fig. 1a.}
\begin{center}
\includegraphics[angle=270, width=5in]{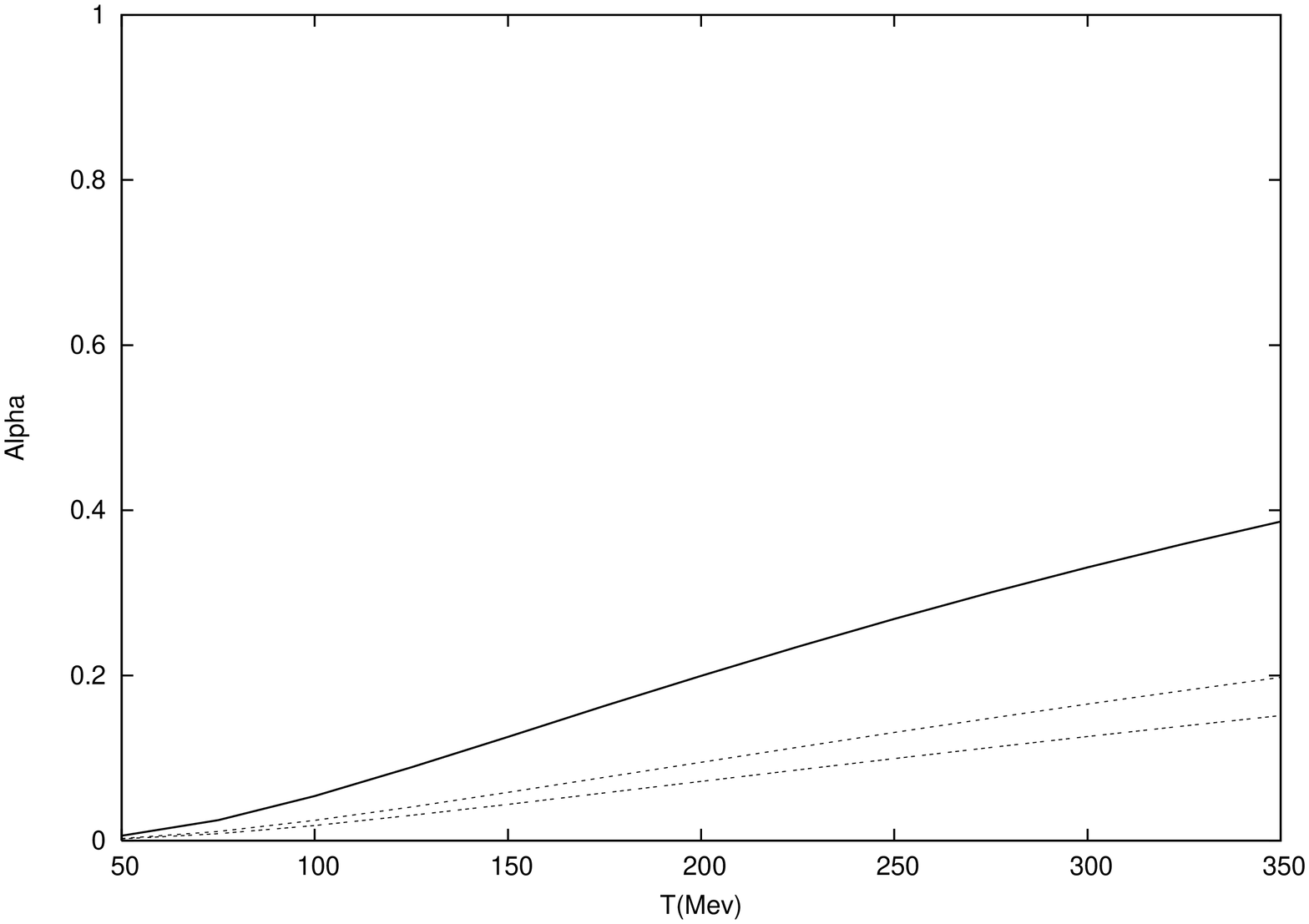}
\end{center}
\end{figure}

\end{document}